# HMIoT: A New Healthcare Model Based on Internet of Things

Mohsen Yaghoubi Suraki [1], Morteza Yaghoubi Suraki [2], Leila SourakiAzad [3]

[1] Department of IT and Computer Engineering, Islamic Azad University, Qazvin Branch
Qazvin, Iran

[2] Department of Human Science, Islamic Azad University, Sari Branch
Sari, Iran

[3] Department of Human Science, Islamic Azad University, Chalus Branch
Chalus, Iran

**Abstract**
In recent century, with developing of equipment, using of the internet and things connected to the internet is growing. Therefore, the need for informing in the process of expanding the scope of its application is very necessary and important. These days, using intelligent and autonomous devices in our daily lives has become commonplace and the Internet is the most important part of the relationship between these tools and even at close distances also. Things connected to the Internet that are currently in use and can be inclusive of all the sciences as a step to develop and coordinate of them. In this paper we investigate application and using of Internet of things from the perspective of various sciences. We show that how this phenomenon can influence on future health of people.

## 1. Introduction

With the development of information technology and its rapid entry into the daily lives of people, information has gained an important Special position. Today, accessing to information is not only as a necessity but also it's as a powerful tool for controlling our world. Using social network websites such as Facebook and Twitter, existing internet filtering in developing country, covering online news, Smart TV connected to the internet and exposed information in the internet without any restrictions lead governments to have a good and accurate plans for putting their policy to their countries. On the other side, widespread access to the Internet has allowed people to communicate with each other easily in far distance and even find a job or if they become ill they find and go to the nearest clinic. So In the new century, three things will play an important role in people's lives: 1. The information 2. Internet 3. Things connected to the Internet. in the future, these parts with increasing quantity and improving quality and with rapid developing of Internet-connected devices have direct effects on people's life.

In 2005, the population used the Internet was reached to 1 billion people and in 2010 it was reached to 2 billion people. In 2014, these value reaches to 3 billion people. This process is illustrated in the following diagram [1]. Based on [2] network at 2017-2020 include 7 trillion wireless telecommunication units for 7 billion people and total number of things that could be connected with the future network is 50 trillion[3].

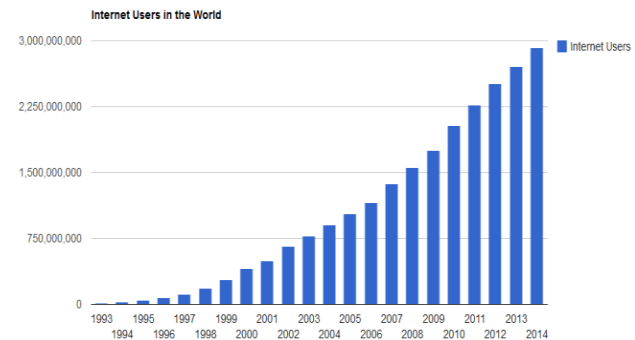

Fig.1 Trends of internet users [1]

The Internet is used by many peoples around the world for accessing to many resources such as play online games, make social networking or other information we need. Moreover, the Internet is playing a global platform to interconnect physical objects or ''things'' and virtual objects. In fact enabling new ways of working, interacting, entertaining, and living is expected from new viewpoint of internet [4-5].

According to this prediction, not only physically large number of connected objects will grow, but also in term of number of users and each user's usage of those objects will increase. The widespread applications is not limited to one part of the world, it will also include a broad range of various science.

## 2. Internet of Things (IoT)

In the last decade, a novel paradigm named internet of things (IoT) was popularized by the work of the Auto-ID Center at the Massachusetts Institute of Technology (MIT) [6]. Internet of Things breaks traditional thinking mode, connects physical infrastructure with IT infrastructure, and lets all the things connect with the network [7]. There are many key parts for defining IoT concept in which emphasis on internet ability, structural features, RFID tags and using smart ability of them. According to International Telecommunication Union (ITU), the definition of the IoT is "from anytime, anyplace connectivity for anyone, we will now have connectivity for anything" [8]. Indeed, Internet of Things (IoT) is an integrated part of Future Internet with self-configuring capabilities based on standard and interoperable communication protocols where physical and virtual "things" have identities, physical attributes, and virtual personalities [9]. As shown in left side of fig.1 The research roadmap from the European Commission (2009) shows that how interaction among the real physical, the digital, virtual worlds and society will make IoT as a tool for integrating knowledge [10].

2.1

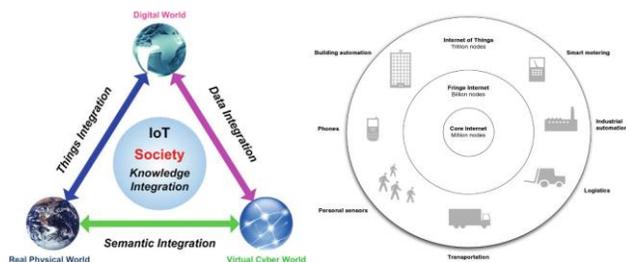

Fig 2. Interaction among the real physical, the digital, virtual worlds [10-11]

In right side of figure 2, Core internet of backbone routers and servers, including millions of nodes (any kind of network device) in total. The equipment connected to the internet includes all the personal computers, laptops and local network infrastructure. Internet of things encompasses all the embedded device and networks that are natively IP-enabled and internet-connected, along with that internet service monitoring and controlling those devices [12]. Things will configure themselves with "intelligent/cognitive" behavior in their new environment, when faced with other things and unforeseen circumstances [12]. According to [13] the IoT has following domains:

- Transportation and logistics domain.
- Healthcare domain.
- Smart environment (home, office, plant) domain.
- Personal and social domain

There is another concept called Internet of everything (IoE) that is located on internet of things. It means IoE can realize with IoT concept. According to [14] IoE is capable of helping organizations achieve many public-policy goals, including increased economic growth and improvements in environmental sustainability, public safety and security, delivery of government services, and productivity. Due to widespread using of IoE, we just focus on IoT's Definitions, Concepts and applications.

Although IoT widely used to define connected everyday objects, the nature of the connection should be determined. According to [15] two distinct modes of communication in the Internet of Things are specified:

- Thing-to-person (and person–to-thing) communications focuses on technologies and applications wherein people interact with things and vice versa. For example having remote access to objects by humans such as access to CCTV camera objects in hospital to control patients, remote environmental protection for rare animals, forest fires and so on. these objects continuously report their status, whereabouts, and sensor data.

- Thing-to-thing communications focuses on technologies and applications wherein everyday objects and infrastructure interact with no human originator, recipient, or intermediary. Machine-to-machine communication is a subset of thing-to-thing communication; but it often exists within large-scale IT systems and so may not consider as "everyday objects".

In healthcare domain. According to [16] we have another categorization of communication in IoT that focus on healthcare system. In fact, it's subset of mentioned categorization that shows detail of component in healthcare system. It also shows that how dynamic capabilities of Internet of things can use in that system:

- D2M (Device-to-Machine)
- O2O (Object-to-Object)
- P2D (Patient-to-Doctor)
- P2M (Patient-to-Machine)
- D2M (Doctor-to-Machine)
- S2M (Sensor-to-Mobile)
- M2H (Mobile-to-Human)
- T2R (Tag-to-Reader)

In the structural sense, Internet of things has three basic components: RFID systems, middleware systems and Internet systems. RFID system is the wireless non-contact equipment which leads to transfer data in order to automatically identifying and tracking tags attached to objects and according to particular purpose, RFID reader processed data from portable devices named tag [17]. The data transmitted by the tag provides online monitoring of patient's location information, patient's physical information such as age, sex, blood pressure, glucose level or nearest nurse for emergency events. With RFID systems, we can monitor healthcare objects in real-time, without the need of being in line-of-sight. It means that we can map real world healthcare system into the virtual world system. Middleware savant system is software that located between RFID hardware and healthcare applications [18].

## 3. IoT Applications in Science

In recent years, in the various sciences we are witness to grow science in many fields. Advancement of computer technologies leads information to exchange quickly and more accurately. Exchanging this information requires secure and accessible equipment to transfer. Internet of Things in its conceptual meaning able to combine various sciences. In this section, basically we define and categorize various sciences and then we show that how is IoT framework for different science and how IoT can effect on changing science in an advanced way.

Science has many different definitions, based on [19] it defines as the pursuit and application of knowledge and understanding of the world following a systematic methodology is based on the evidence. On the other hand, according to Merriam online dictionary [20] as knowledge about or study of the natural world is based on the facts obtained by experiment and observation. The other definition of this dictionary refers to a particular field of study. According to the Longman Dictionary [21] it is knowledge about the world based on examining, testing and proving the facts. According to the Oxford Dictionary [22] it defines as intellectual and practical activity involves systematic study of the structure and behavior of the physical and natural world through observation and experimentation.

In all mentioned definition, knowledge is an important part of science and for reaching to knowledge we need to have systematic study and true experiences and accurate documentation about the natural words. As we know today our virtual world is growing the same as our natural world. Internet will connect everyplace in near future. According to [23] Google are going to spend more than $1bn on a fleet of 180 satellites to beam internet access to unconnected parts of the globe. It means we should consider capacity of our virtual world because of

intelligent computing with super computers, providing accurate information and documentation and finally having unlimited capacity for saving that information. In the past, a piece of paper has play a lot of role but today making intelligent objects such as smart phone, smart TV, intelligent equipment and finally online communication between them lead to change our new world. Therefore, they have a major contribution to the advancement of science.

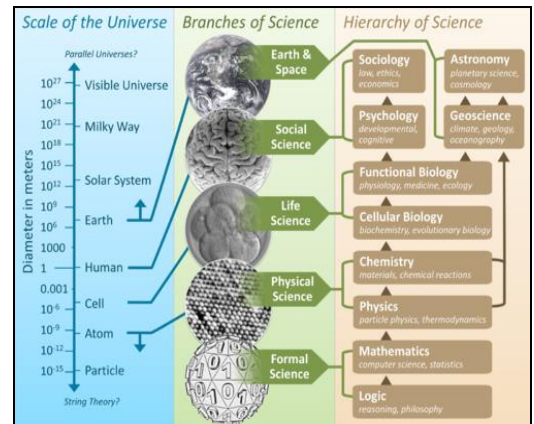

Fig 3. A perspective for various science [46]

As shown in this figure 3, science is divided into five categorize (Earth & Space, Social Science, Life Science, Physical Science and formal Science) that each categorization has own branches (Astronomy, Psychology, Chemistry, Mathematics and so on) [46]. In each science, there are inseparable intelligent objects. Communicating these object based on internet, we can sure that processing on information will perform better than past. It means an object with enough intelligent and permanent online communication can perform many instructions in real-time. Internet of things concept plus science model can communicate together as shown in figure 4. For example, in spacecraft astronaut's cloth, ring, or their other own objects can communicate together and based on its local knowledge react to many situations. In another example in psychology, Internet of things can use in physical problems like disabilities problems [34] or in controlling mental disorders like depression [24], obsessive-compulsive [25]. Because objects with having ability to communicate with each other, they can perform permanent monitoring on their determined patients and before the occurrence happen or upon the occurrence of a traumatic event some notifications send to nearest medical center for informing nurses to what happened for their patients.

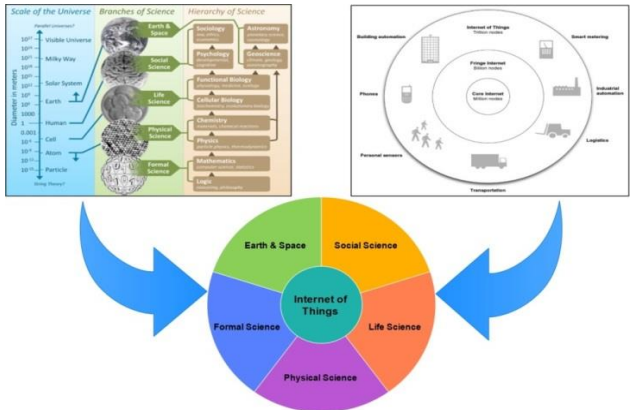

Fig 4. Relationship between Science based on IoT

In psychology viewpoint, depressed people are usually tired and do not have the patience and energy, resulting in much less use of objects such as mirror for their appearance. These people use very little iron and may reduce their use of Television and telephone or to be at risk of obesity. They are more or less sleep than usual; resulting in the use of beds in them is different from ordinary people. Due to the emergence of depression, buying food and clothe will be decreased in addition the amount of useful work in their workplace and their business transaction are low. People's information can be stored in local database by using the internet of things and on the information collected to identify whether they are depressed or at risk of depression or are living in a normal and healthy. For OCD patients, they may have repetitive behaviors such as repetitive washing hands or dishes, repetitive cleaning table and chairs, or repetitive extreme symmetrical arranging of things. Using Internet of things technology and environment intelligence, the number of times things are used in a house can be obtained and this information can be made available to the therapist.

In chemistry, researchers in laboratory can easily work on their tools because of secure and intelligent objects. It means glass mug (objects) equipped with sensors that connected to internet and other objects can help people and researchers to preserve their safety when they are working with hazardous material. It also can give some ingredient information that exists in mug. This information can show the result of research or show how much material mixture. In the other area intelligent object can help people such as learning children or growing culture level of society. For example, imagine our child wants to go to school but he/she forgets to carry pencil or notebook. With an intelligent bag, before our child goes to school he/she can notice what's forgot before leave home. Bag can use other object in house like smart TV or can use own screen to inform children or parents. Teachers can also notify to their children remotely if they need children to carry specific object for next class. By this process not only relations between children, parents, and teachers will become better but also Educational process will improved.

## 4. Healthcare Models

According to [26] The World Health Organization defines health system as follows: "A health system consists of all organizations, people and actions whose primary intent is to promote, restore or maintain health. This includes efforts to influence determinants of health as well as more direct health-improving activities". It means that how we can afford caring for our patient like child, sick or elderly people, which facilities can deliver personal health services. it shows that we need real-time activities in terms of distance limitations. IoT makes a process to easily access patient information and online monitoring for them.

There are many models for healthcare, but all of them use the same steps for modeling their systems. As we focus on healthcare system, National Institute for Children's Health Quality describes a model for child that formed into six major parts [27]. As show in fig 5, they categorize these models into 6 steps:

1. Healthcare System and Organization (for receiving in individual practice settings and the practice's organizational setting and policies),
2. Community Resources (Day care, Head Start, schools, and after-school programs are just some of the essential community resources that must be integrated with health care),
3. Family and Self-Management Support (ability of children and families to manage their own care, focusing on the family's role in managing their child's well-being and illness, collaborating with families in setting shared goals for child and providing educational materials and resources to support them),
4. Delivery System Design (need of existing a multidisciplinary team with clear understanding of roles and how each contributes to a child's care, Team members should have sufficient training for their roles and should communicate together often),
5. Decision Support (Practices should embrace evidence-based guidelines, These guidelines should be embedded in documentation systems, Primary care practitioners should have access to specialty expertise from these document and specialty consultation and supervision)
6. Clinical Information Systems (Information technology can be used to identify entire populations of children with specific needs, assess practice performance, target high-risk populations, and plan for future needs).

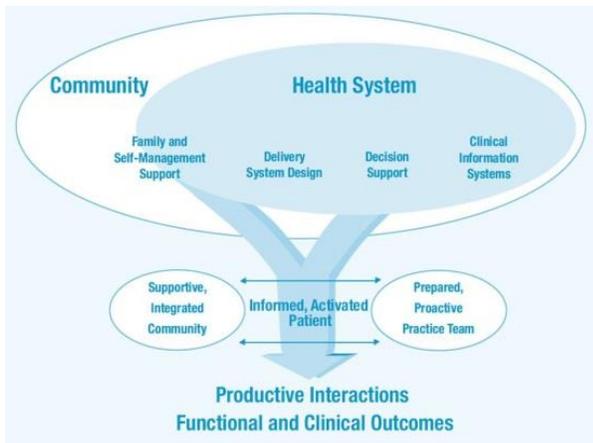

Fig 5. A Healthcare Model [27]

Based on the global recession and to reduce the rising healthcare costs associated with chronic diseases, Health Research Institute has identified an emerging phenomenon in every major health system around the world. Government and health industry leaders should consider fundamental structural changes in the health system to help the people. With the advancement of technology and the development of communication tools and with health care organizations assistant, We expected that the healthcare models change in term of efficiency and effectiveness and virtual care will use in order to facilitate information sharing and to provide interactive between things and human [28].

## 5. Healthcare Model Based on Internet of Things (HMIoT)

The cost of healthcare is increasing and the need for alternative methods that can be economically affordable and permanent monitoring of the patients is necessary. As we mentioned, intelligent object not only eliminate the needs of clients and most people but also they can act as doctor or teacher for controlling or monitoring them remotely. This needs do not limit in a particular time or a particular place. It wildly uses in our physical and virtual world. It means that objects naturally move from bottom-level concept to an abstract-level concept in passing time because of changing their role in our society. We can apply this intelligent role in society. Integrating knowledge database between them and online connecting together is a complex work so that it is necessary for planning for government's policies. As shown in figure 6 authors in [29] consider IoT as abstract levels and they divide into three main level.

**Public Level:** On a public level, such as air pollution, humidity and food shops as well as the amount of traffic in such things like cars and people lined up to take the bus or in the street, governments can get these useful information to choose a good policy for managing their city. Some papers on this area are shown in first column of table 1.

**Personal Level:** On a personal level such as elderly people or place of people or expiration dates the materials in the refrigerator, we can use IoT and other Technologies like WBANs (Wireless Body Are Networks), NFC, Bluetooth together. Some papers on this area are shown in second column of table 1.

**Object Level:** In object level, object's status will be considered. If object is going to fail, we will plan to exchange it with a health object. Some papers in this area are shown in third column of table 1.

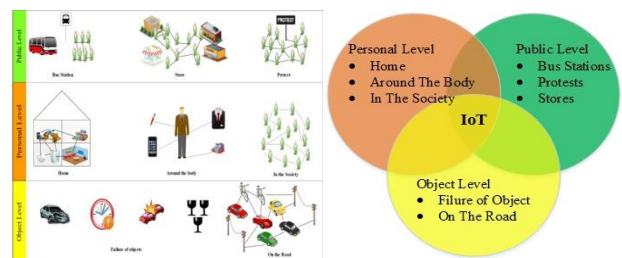

Fig 6. Three major level of IoT in Public Health Approach [29]

These three level live in our hide life and they have direct effects in our life. They have unlimited capacity in term of human capacity. Internet of things with goals that are defined by application layer and the type of application opens a new way to improve everyday activities. There are a lot of example in public level that help people and government to control what they want. The objects where located in the city and with the people that connected to the internet, can help the elder people for air pollution with gathering information of objects about the air. Additionally, People can find out when the bus will depart and how many people are in the queue of the bus station or how many people are in the traffic. On the other hands, governments get the numbers of people who are participated in a protest and/or how many tourists are in specific areas. Meanwhile, by gathering information from objects we can find out whether there is any parking place for our car. Another example, by using it we can perceive the expiration dates of the foods in refrigerator or getting information and price of stuffs in mall before going there[29].

Table 1. classification of some major papers into IoT Approach

|  | Public Level | Personal Level | Object Level |
|---|---|---|---|
| [16], [32-33], [34] | ✓ |  |  |
| [16], [30-38] |  | ✓ |  |
| [39-45] |  |  | ✓ |

We describe how much object's treatment can effect on healthcare for people. Now we ask a question which level is important to notice? As shown in figure 7, the importance of preservation of objects is from object level to personal level and the public level. Because an object can help to healthcare system in personal and public level when objects be health in object level.

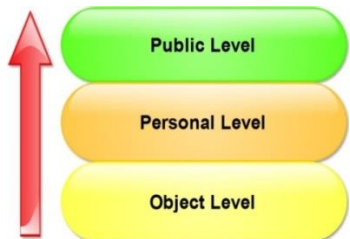

Fig 7. Bottom-up importance of components in Public Health Approach

As shown in figure 8, we extend models two above mention model into one general model called Healthcare Model based on IoT (HMIoT). Because of existing important role in objects for healthcare system, we extend and improve this model with combining three level of IoT into healthcare model. This model improved social health level and decrease cost of healthcare. In some times, object show happens clearly in term of other tools. For example Some NGOs or reporters want to know exactly how many people attend in particular protests or may they want to know how many things failed or damaged. It shows that how much the protest was violent. In many case, This model can show accurate information to relevant organization without any censoring. In another example, if a person injured in an car accident and her/him car was damage completely, relevant organizations can understand and predict how is person's situation and how much they need to help. It also shows the place of accident. This information is very important for helping people when they fall into bad situation. Imagine a person injured and we don't know where event is occurred. If emergency centers unable to predict the place of event, One human's life may be compromised. In the basic healthcare system model, objects and its effects on accurate information has been less attention, but in this model we consider objects as main part of healthcare and we have been attention not only on people's health but also on object's health.

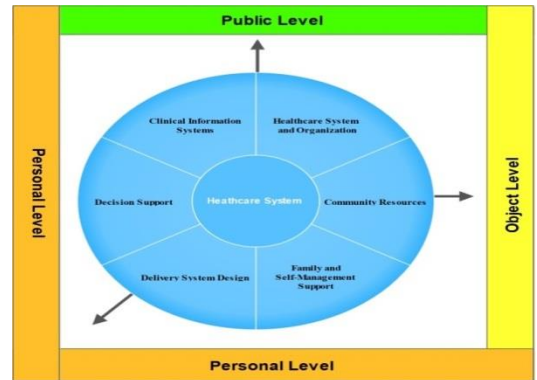

Fig 8. Healthcare Model based on IoT (HMIoT)

Existence of three public, personal, object level for healthcare system is necessary. Every objects can determine which applications they use and which type of them we need for gathering information. In healthcare system, we can determine position of each application. For example, in public health approach, the object and public level has more usage than personal level for getting social resources or in family support domain and self-management objects and personal level more useful than public level. In this model, making policies will become complex and need for validating of them will be increasing because of considering both people health and object health. As we mentioned, objects are unseprated part of our life and if they want to have direct effects on our life we need to use them integrately and do not limit to specific area. For example, if our foods in refrigrator expires, it's need someone who use that food know about it. Imagine healthcare system perform some actions after happen accure but in this situation previouse model can not help that person because him/her life may be compromised. Our model lead to use accurate information and rapid act in term of previous healthcare model. As an example in the police office, many people lost because of murdering, being strange and forgetting address. The objects with unique identification such as smart watch or smart phone or smart shoes can easily inform th policeman to find lost peoples. This siuation can even prevent terrorists attach before they happen.

## 6. Conclusion

In recent years, increasing pupulation and medical costs people can not be able to pay their healthcare. Designing and modeling an efficient system to increase healthiness leven in society is critical. Utilizing intelligent objects in IoT that located in our inviroment can help us to design a new healthcare model. Objects connected to the internet can chage basic healthcare model. Therefore, it can not only improve public health of our society, but also decrease medical and healthcare costs for most of patients. We even use the IoT concept in othe sience that is need to be changed. However in this paper we don't focus on some details such as security, layer and cross layer structures but we introduce a new model named HMIoT and open a way for researchers to focus on increasing healthcare level. Paying attention to public health and improving human life are neccecary part of our life that sometimes we don't see them. So we need to have an careful plan for improving them.

**Mohsen Yaghoubi Suraki** received his Bachelor`s degree in Information Technology from Qaemshahr Islamic Azad University in Iran in 2011. He received his master's Degree in Information Technology from Qazvin Islamic University in Iran in 2014. He is also lecturer of some university colleges in Iran. He has many papers about using IoT in healthcare, mental disorders, and about NFC and MANETs. His current research interests include internet of things in healthcare system, psychological disorders and computer networks.

**Morteza Yaghoubi Suraki** received his Master`s degree in general psychology from Sari Islamic Azad University in Iran in 2012. He has many international and national papers about mental problems and using new technology in psychology. His current research interests include internet of things in healthcare System, psychological disorders, mental problems and cognitive therapy.

**Leila SourakiAzad** received her Bachelor`s degree in Primary Education from Sari Islamic Azad University in Iran in 2010. She is currently master student of Curriculum Planning in Chalus Islamic Azad University. She is also a teacher in primary school. Her current research interests include child and adolescent mental health, educational models and skills training for children.